\documentstyle[11pt,newpasp,twoside,epsf]{article}
\def\edcomment#1{\iffalse\marginpar{\raggedright\sl#1\/}\else\relax\fi}
\reversemarginpar

\newcommand{\msun}{{\rm M_{\odot}}}

\begin{document}
\title{A New Evolutionary Picture for CVs and LMXBs}

\author{A.~R. King and K. Schenker}

\affil{Theoretical Astrophysics Group, University of Leicester, 
  Leicester, \mbox{LE1 7RH}, U.K.}


\begin{abstract}
We consider an alternative to the standard picture of CV and LMXB
evolution, namely the idea that most CVs (and by extension LMXBs) may
not yet have had time to evolve to their theoretical minimum orbital
periods. We call this the Binary Age Postulate (BAP). The observed
short--period cutoff in the CV histogram emerges naturally as the
shortest period yet reached in the age of the Galaxy, while the
post--minimum--period space density problem is removed. The idea has
similar desirable consequences for LMXBs. In both cases systems with
nuclear--evolved secondary stars form a prominent part of the
short--period distributions. Properties such as the existence and
nature of ultrashort--period systems, and the spread in mass transfer
rates at a given orbital period, are naturally reproduced.

\end{abstract}


\section{Introduction}

The current picture of CV evolution (see e.g. King, 1988 for a review)
has remained essentially unchanged for more than two decades. Its main
elements are the propositions that

{\it 1. Formation:} Common--envelope (CE) evolution produces pre--CVs
   close to contact at all periods $P$ such that $1.5 \la P \la 12$~hr

{\it 2. Secular evolution:} angular momentum loss brings pre--CVs
   rapidly into contact after the CE phase and also drives their
   subsequent evolution as CVs, accounting for the main features of
   the observed CV period histogram. The loss mechanism is
   gravitational radiation (GR) and some rather stronger agency,
   perhaps magnetic braking (MB) at periods $P \ga 3$~hr

{\it 3. Youth 1:} CVs are significantly younger than the Galaxy, so many
   generations of CVs have passed through the observed CV period range

{\it 4. Youth 2:} CV secondaries are not nuclear--evolved

The great difficulty inherent in theoretical studies of CE evolution
via numerical simulations has meant that elements 1, 3 and 4 have
remained largely unexamined. Until recently, most researchers have
concentrated on tying down the details of 2.\ above.

The evolution of low--mass X--ray binaries (LMXBs) is more complex
than that of CVs because of the need to form a neutron star or black
hole through a supernova explosion, with a resulting potential for
disruption of the binary. In particular proposition 1.\ is no longer
obvious. Nevertheless the similarity of {\it short--period} LMXBs to
CVs has encouraged tacit adoption of at least 2.\ and 4.\ for them
also. Significant differences (cf Fig.\ 1) in the period histograms
of the two types of binary (first pointed out by White \& Mason, 1985)
have received relatively little attention.

Here we suggest that for CVs there are real reasons to question the
three linked propositions 1.--3.--4., and even some of the apparent
successes of 2. We suggest a way out of these difficulties which
appears promising for both CVs and LMXBs.


\begin{figure}
\plotfiddle{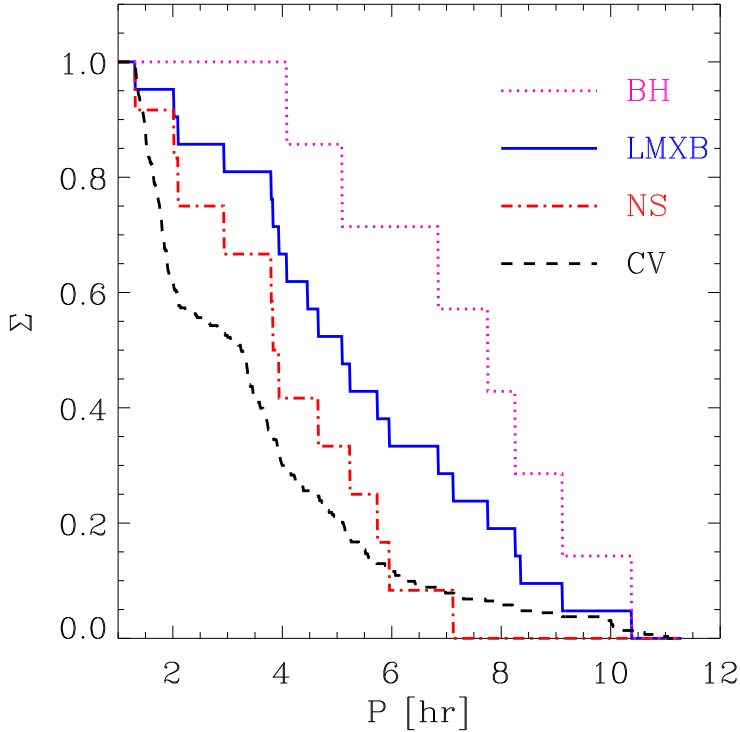}{9cm}{0}{75}{75}{-175}{-30}
\caption{Comparison of the LMXB period distribution (cumulative only)
with that of CVs (dashed) over a typical range of orbital periods.
Data are taken from Ritter \& Kolb (1998, with updates). Additional
curves are shown for LMXBs with established NS (dotted) and BH
accretors (dash-dotted). Note the systematic trend between
different primaries.
}
\end{figure}

\section{The CV period minimum as an age effect}

The observed CV period histogram (Fig.\ 2) cuts off sharply at an orbital
period of $P = P_0 \simeq 76$~min. There are respectively 0 and 12
systems in the period ranges $P_0 \pm 5$~min. The idea that $P_0$
represents a global period minimum ($\dot P = 0$ for $P_{\rm min}$) for
CVs has been widely accepted for the last two decades. It is clear that
such a global minimum can exist (Paczy\'nski \& Sienkiewicz, 1981;
Paczy\'nski, 1981; Rappaport, Joss \& Webbink, 1982). As the mass $M_2$ of
an unevolved secondary star in a CV is reduced by mass transfer, the
binary period $P$ decreases also. However for very small 
$M_2 \la 0.1 \, \msun$, the secondary's Kelvin--Helmholtz time 
$t_{\rm KH}$ begins to exceed the timescale 
$t_{\rm M} = -M_2/\dot M_2$ for mass loss driven by angular momentum
loss, e.g. by gravitational radiation. 
At this point the star will expand adiabatically, causing the binary
period to increase rather than decrease.

\begin{figure}
\plotfiddle{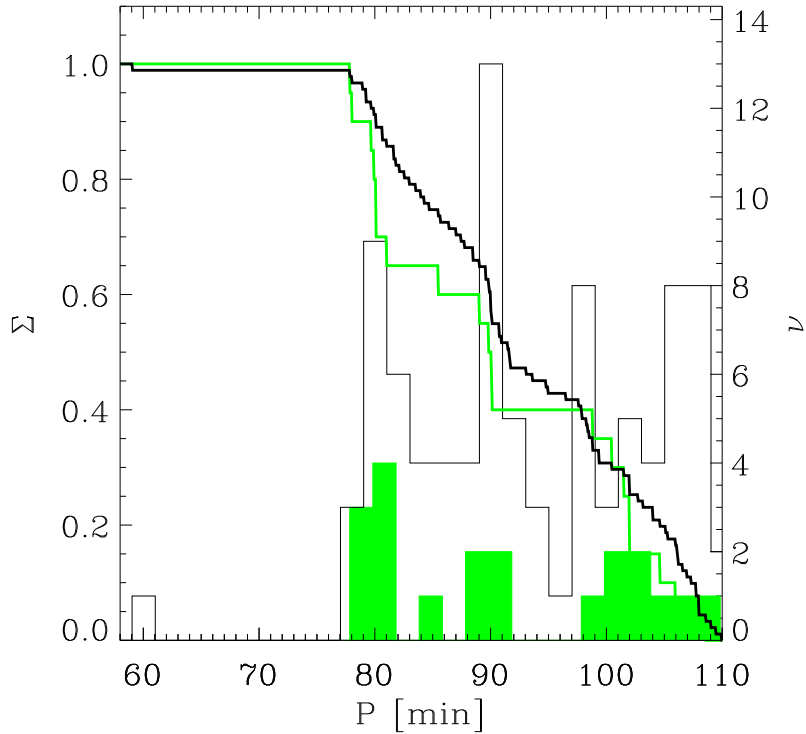}{9cm}{0}{75}{75}{-175}{-30}
\caption{Observed CV period distribution (binned and cumulative) at
periods around $P_0$. Data are taken from Ritter \& Kolb (1998).
Overplotted in grey is the subclass of AM Her systems showing
essentially identical behaviour. The single system at 
$P \simeq 60 \, {\rm min}$ is V485 Cen.}
\end{figure}
Detailed calculations always predict a value $P_{\rm min}$ very close to,
if slightly shorter than, the observed $P_0$. The discrepancy 
$P_{\rm min} < P_0$ is persistent, but may reflect uncertain or
over--simple input physics (cf Kolb \& Baraffe, 1999). However there
is a much more serious problem with this interpretation of the
observed cutoff at $P_0$. This concerns the discovery probability 
\begin{equation}
p(P) \propto {(-\dot M_2)^{\alpha}\over |\dot P|}.
\label{prob}
\end{equation}

Here $\alpha$ is some (presumably positive) power describing
observational selection effects (e.g. $\alpha = 3/2$ for a bolometric
flux--limited sample). Since $\dot P = 0$ at $P = P_{\rm min}$,
$p(P)$ must clearly have a significant maximum there unless 
$-\dot M_2$ declines very sharply near this period. In other words, the
observed CV period histogram should show a sharp rise near a global
minimum $P_{\rm min}$ unless the mass transfer rate drops
there. However all evolutionary calculations show that $-\dot M_2$
changes very little as $P_{\rm min}$ is approached. We conclude that
there should be a large `spike' in the CV period histogram near a
global minimum $P_{\rm min}$ (cf Kolb \& Baraffe, 1999).

The lack of such a spike in the observed period histogram (Fig.\ 2) has
prompted numerous theoretical investigations. Many of these propose ways
in which CVs might become difficult to discover near $P_{\rm min}$. A
basic problem for this type of argument is that, as we have seen,
there is nothing at all unusual about the system parameters (mass
transfer rate, separation etc) at this period. Further, attempts to
use accretion disc properties as a way of making systems hard to
discover founder on the fact that the AM~Herculis systems, which have
no accretion discs, have precisely the same observed short--period
cutoff $P_0 \simeq 80$~min, and no spike either (cf Fig.\ 2).

The identification of the observed cutoff $P_0$ with the global
minimum period $P_{\rm min}$ creates a second problem, particularly
emphasized by Patterson (see the review in this volume, and references
therein). Namely, if many generations of CVs have completed their
evolution and passed the minimum period in the history of the Galaxy,
the predicted space density of post--minimum CVs becomes uncomfortably
high. The nearest systems should be close enough to be detectable even
as bare white dwarfs; and the problem gets worse when one realises
that mass transfer decreases only very slowly (timescales $\sim
10^{10}$~yr) after passing $P_{\rm min}$, so that they are definitely
brighter than the bare white dwarfs. Since the orbital period also
changes very little, the result should be a very large number of
nearby CVs with brightness and periods very close to those at $P_0$,
which are not observed.

In view of these and other difficulties, it seems reasonable to consider
dropping the assumed identification of the observed cutoff $P_0$ with the
global minimum $P_{\rm min}$. Thus, from now on we will instead
investigate the idea that $P_{\rm min}$ might be genuinely shorter than
$P_0$, or more succinctly, {\it that even the oldest CVs have not yet
reached $P_{\rm min}$.}

The timescale $t_{\rm evol}$ for the secular evolution of CVs down to
$P_{\rm min}$ is considerably shorter than the age of the Galaxy, even
from the longest commonly observed periods $\sim 8-10$~hr (assuming
that magnetic braking is not drastically reduced as has been recently
proposed -- see the article by Pinsonneault in this volume). Thus to
maintain the idea that $P_{\rm min} < P_0$, we must require that most
CVs came into contact only a time $< t_{\rm evol}$ ago. In the
conventional picture where CE evolution produces pre--CVs close to
contact at all orbital periods, this would mean that CVs emerged from
CE evolution relatively recently in the age of the Galaxy, presumably
as the result of a starburst. This seems unlikely, so we assume
instead that for most systems {\it the time $t_{contact}$ to shrink the
binary enough to initiate mass transfer is at least comparable to the
Galactic age $t_{\rm Gal}$} (more precisely $t_{contact} \ga t_{\rm
Gal} - t_{\rm evol}$).

We call this idea the Binary Age Postulate (BAP). In the usual
language, it amounts to requiring {\it either} that CE evolution is
more efficient in removing the envelope of the white dwarf progenitor,
thus leaving wider systems than usually assumed, {\it or} that with
conventional CE evolution, orbital decay into contact by angular
momentum loss is much slower than usually assumed. The second
possibility would fit with the idea of drastically reduced magnetic
braking (see the article by Pinsonneault in this volume). However,
unless the usual value of the magnetic braking torque is restored once
the system reaches contact there are obvious problems in explaining
the brighter (novalike) CVs and the period gap itself. Hence while the
reduced braking idea is worth bearing in mind, for the expository
purposes of this paper we shall assume the first possibility,
i.e. that BAP is satisfied because CE evolution is more efficient
than usually assumed. Thus most CVs with secondaries massive enough to
have magnetic braking would emerge from CE evolution with periods of
order 12~hr or more. CVs with lower--mass secondaries could emerge
with shorter orbital periods, but such that relatively few reached
contact (evolving via gravitational radiation) within $t_{\rm
Gal}$. We note that this type of distribution is not in conflict with
the observed pre--CV distribution (Ritter \& Kolb, 1998).

Armed with these assumptions we can give two immediate consequences
for CVs, and two for LMXBs.

(i) {\it The period distribution near $P_0$}. The characteristic square
shape of this distribution is naturally reproduced, provided only that CVs
at longer periods decrease their periods more quickly than those at short
periods, e.g. $-\dot P = G(P)$ with ${\rm d}G/{\rm d}P > 0$. This is of
course true for the usually assumed forms of magnetic braking. Hence the
observed distribution appears naturally if CVs generally come into contact
with secondary masses $M_2$ large enough ($M_2 \ga 0.3 \, \msun$) for their
pre--contact evolution to have been driven by this mechanism. This is
of course precisely the content of the BAP idea.

(ii) {\it The space density problem.} This is removed, since there is
no presumption that many generations of CVs have passed the observed
cutoff $P_0$, and thus no presumed rate at which CVs are piling up in
the Galaxy. Assuming that the formation rate of pre--CVs has decreased
markedly since the early epochs of the Galaxy, there is no
corresponding problem with the space density of pre--CVs. Note that if
we do not make the latter assumption, the space density problem is
inevitable in {\it any} picture: CVs pile up either as post--minimum
or pre--contact systems, as we presumably cannot destroy the white
dwarfs in either state.

(iii) {\it The LMXB minimum period.} Magnetic braking acts more slowly
on LMXBs than CVs, as the binary inertia is greater. Thus we might
expect LMXBs to have longer $P_0$ cutoffs than CVs -- consistent with
observation (see Fig.\ 1) -- as they will presumably be unable to reach
such short periods in the age of the Galaxy.

(iv) {\it The faint transient problem.} King (2000) points out that
LMXBs with very low mass transfer rates, such as any which
have passed the equivalent of the CV global minimum period,
will be readily observable as faint transients. Although some such faint
systems are observed, the total number in the Galaxy is much too low
compared with the number of `normal' LMXBs to interpret them as
post--minimum systems in the standard picture. Again this is as expected
if most LMXBs have not yet reached the global minimum. We shall see a
further feature of these systems explained in the next section.


\section{Thermal--Timescale and Nuclear Evolution}

\begin{figure}
\plotfiddle{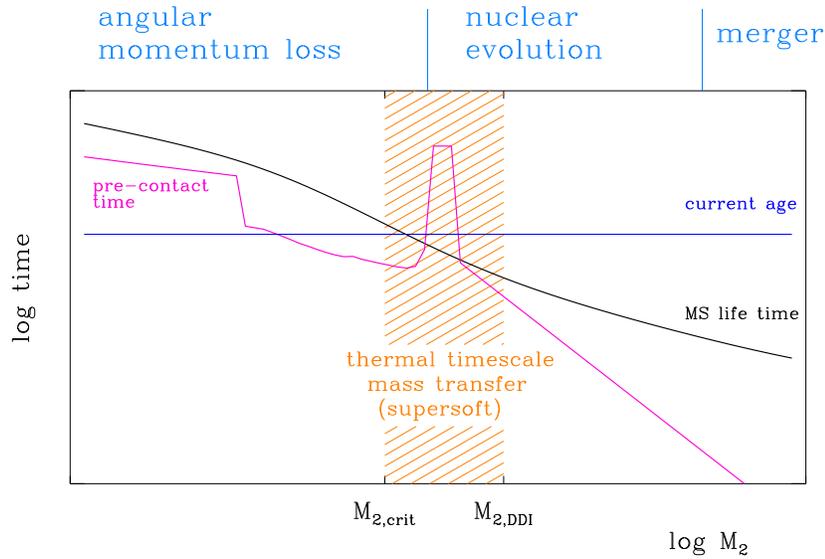}{7cm}{90}{50}{50}{180}{-40}
\caption{Schematic comparison of the duration of pre-contact evolution
with other timescales for close binaries (see text for discussion).}
\end{figure}

\begin{figure}
\plotfiddle{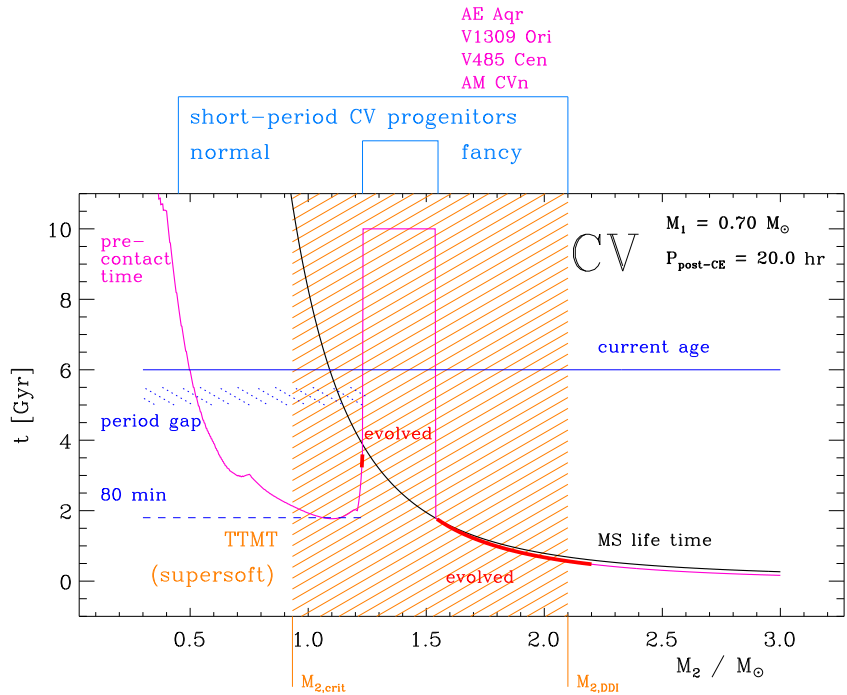}{8cm}{90}{50}{50}{180}{-40}
\caption{Typical CV case within the BAP model: Two progenitor groups
forming the current short-period population of CVs (see text for
details).}
\end{figure}

\begin{figure}
\plotfiddle{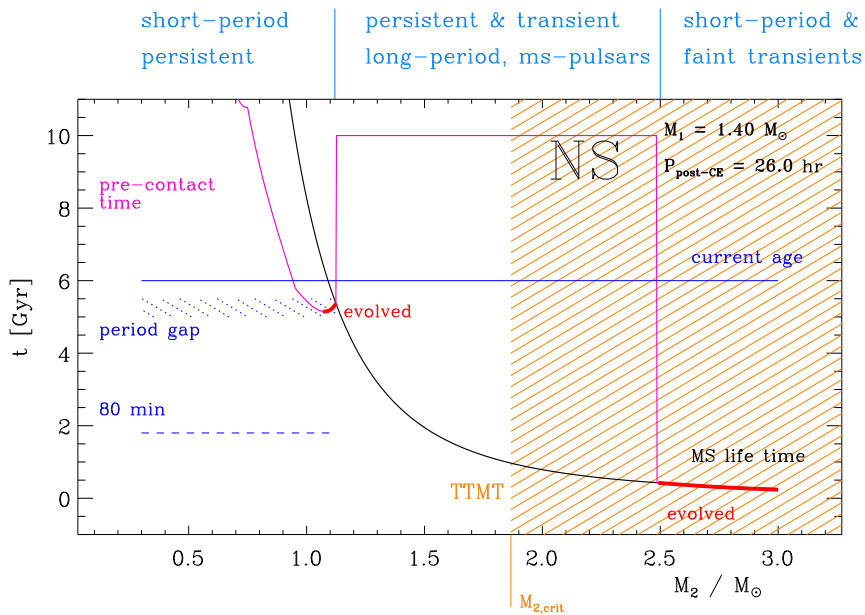}{7cm}{90}{50}{55}{160}{-15}
\caption{Typical NS case within the BAP model: Systems with unevolved
donors are scarce and cannot have evolved below periods around 
$2 .. 3 \, {\rm hr}$ (see text for details).}
\end{figure}

\begin{figure}
\plotfiddle{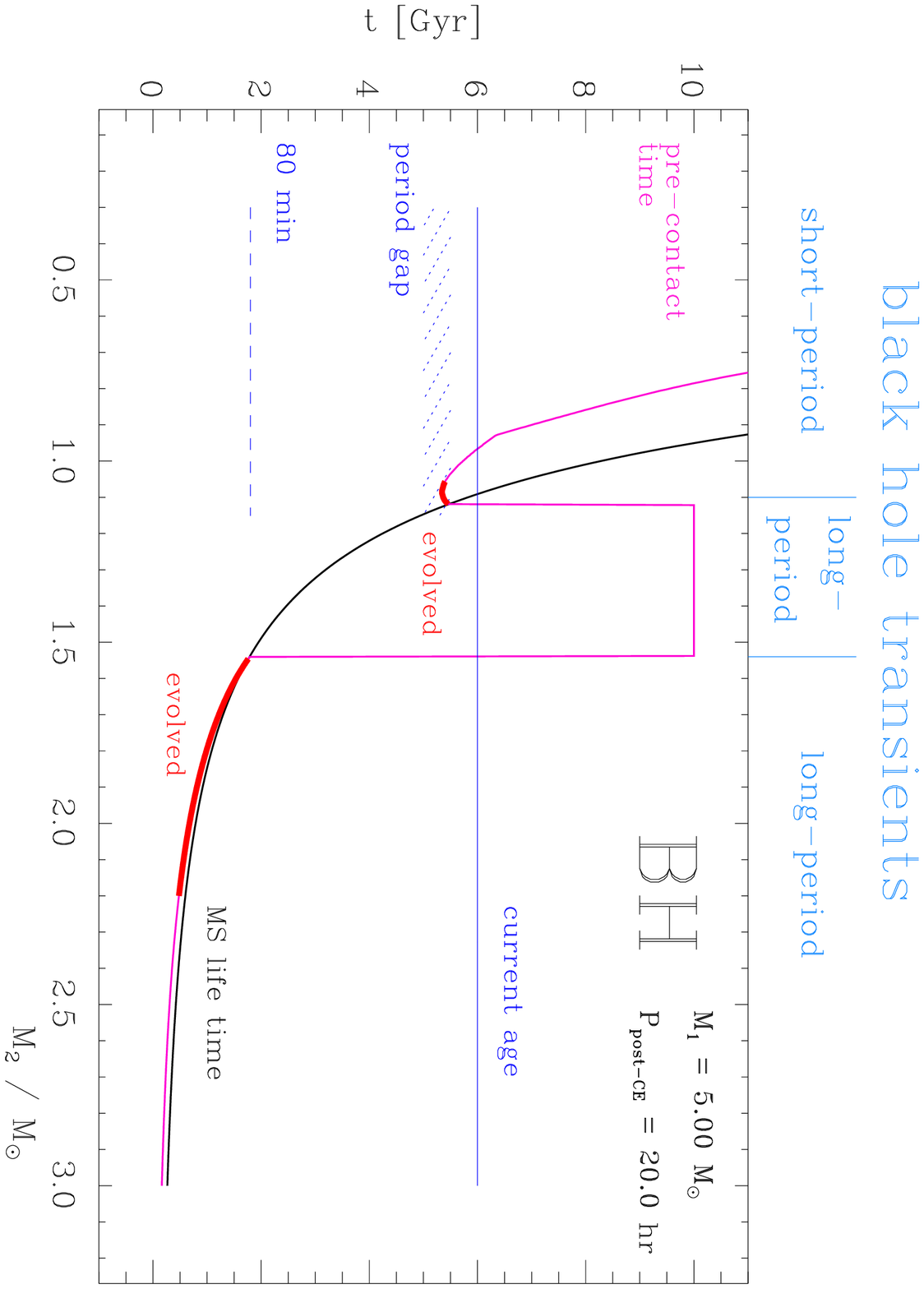}{7cm}{90}{50}{50}{180}{-40}
\caption{Typical BH case within the BAP model: Systems coming into
contact at the higher mass branch do not pass through a TTMT phase and
form long-period LMXBs instead (see text for details).}
\end{figure}

The list (i --iv) above shows that BAP has interesting consequences
for CV and LMXB evolution. Unsurprisingly there are more. The longer
timescales envisaged for orbital decay via angular momentum loss open
the possibility that nuclear evolution of the secondary star might
bring the binary into contact instead, in direct contrast to the older
proposition 4. detailed in the Introduction. This possibility becomes
even more pressing when we allow for the fact that white--dwarf and
neutron--star binaries can survive a phase of thermal--timescale mass
transfer (TTMT) in which $M_2 \ga M_1$ (with $M_1, M_2$ the primary
and secondary masses). For white dwarf systems the TTMT phase, at
least in mild cases with $M_2$ not too large compared with $M_1$, is
probably what drives many supersoft X--ray binaries. The realisation
that some neutron--star systems, notably Cyg X--2, must have survived
quite violent (highly super--Eddington) TTMT is relatively recent
(King \& Ritter, 1999; Podsiadlowski \& Rappaport, 2000; King \&
Begelman, 1999). TTMT may be observable in systems such as SS433, and
the ultraluminous X--ray sources recently identified in external
galaxies (King, Taam \& Begelman, 2000; King et al, 2001).

Figure 3 shows schematically how angular momentum loss and nuclear
evolution may compete in bringing a compact binary into contact. For
low (initial) secondary masses $M_2$ the nuclear evolution timescale
is longer than the age of the Galaxy, so angular momentum loss
automatically dominates. For large $M_2$ nuclear evolution is rapid,
and wins over angular momentum loss. For intermediate secondary masses
thermal--timescale mass transfer may occur if $M_2 \ga M_1$, although
this phase eventually becomes dynamically unstable (the `delayed
dynamical instability') for $M_2$ larger than some critical value
$M_{\rm 2, DDI}$. If some nuclear evolution has already occurred, this
phase can shrink the binary drastically and strip the hydrogen--rich
envelope from the donor, ultimately producing an ultrashort--period
system with a low--mass, hydrogen--poor and probably degenerate
secondary.

Figure 4 shows the situation for the specific case of CVs. The
important result is that BAP hypothesis allows a numerous
population of significantly nuclear--evolved CVs to coexist with the
familiar unevolved CVs envisaged in the standard picture (cf
assumptions 1 -- 4 above). The accompanying paper (Schenker \& King,
this volume) considers this in more detail, and shows that the
resulting distribution has several desirable properties, such as
possibly explaining the spread in mass transfer rates above the CV
period gap.

Figure 5 shows the situation for neutron--star LMXBs. This is
qualitatively similar to the CV case. However the slower orbital decay
here leads to a larger cutoff period $P_0$ (as noted earlier) and a
much stronger tendency to nuclear evolution. Thus many systems evolve
to long orbital periods (days), while as for CVs the short--period
systems include many with significantly evolved secondaries. This
agrees with the deduction by King, Kolb \& Burderi (1996) that the
neutron--star soft X--ray transients observed at such periods must
have nuclear--evolved secondaries, as are now indeed observed in
some cases (e.g. Haswell et al., 2000). As in the CV case,
ultrashort--period neutron--star LMXBs may form after a TTMT phase.

The black--hole case is shown in Figure 6. The larger $M_1$ and
consequently still slower orbital decay intensifies the trends towards
larger $P_0$, more long--period systems, and a greater degree of
chemical evolution in short--period systems noted for neutron
stars. The major difference here is that the smaller mass ratio
$M_2/M_1$ makes a TTMT phase unlikely. As a result very few
ultrashort--period LMXBs with black hole primaries can form, at least
by this channel. If as seems likely the faint transients discussed
above are LMXBs with low--mass highly--evolved secondaries, either
already at ultrashort periods or evolving towards them, this offers a
natural explanation for the observation that almost all of them appear
to contain neutron stars.

A general point emerging from this discussion is that {\it
short--period LMXBs are exceptional}: for most LMXBs nuclear evolution
wins, leading to long orbital periods $\sim 10 - 100$~d. Ironically
these systems spend almost all of their lifetimes as soft X--ray
transients with enormously long recurrence times (cf Ritter \& King,
this volume), and thus remain undiscovered. The greater observational
prominence of short--period systems results from their being either
persistent X--ray sources (neutron--star plus unevolved secondary) or
soft X--ray transients with fairly short recurrence times (all other
short--period LMXBs).


\section{Conclusion}

The BAP idea that even the first generation of CVs and LMXBs have yet
to complete their evolution represents a radical break with the
standard picture of CV and LMXB evolution. However it appears to have
some promising aspects. Given the difficulties with the standard
picture, it seems worthwhile to consider it further.



\acknowledgments{Theoretical astrophysics research at Leicester is
supported by a PPARC rolling grant.}

\end{document}